\documentclass[aps,prl]{revtex4}
\usepackage{graphicx,hyperref}

\begin{document}

\title{Relativity 4-ever?}

\author{Olga~Chashchina}
\email{chashchina.olga@gmail.com}
\affiliation{\'{E}cole Polytechnique, Palaiseau, France}

\author{Zurab~Silagadze}
\email{Z.K.Silagadze@inp.nsk.su}
\affiliation{ Budker Institute of
Nuclear Physics and Novosibirsk State University, Novosibirsk 630
090, Russia }

\begin{abstract}
This essay is about superluminal motion. It is generally believed that special relativity prohibits movements faster than the speed of light. It is explained which motion is actually forbidden by special relativity and why. Tachyons are briefly discussed and it is explained that, due to internal instability, tachyon fields cannot be used to transmit information faster than the speed of light. However, as John Stuart Bell remarked, “what is proved by the impossibility proofs is lack of imagination”. Inspired by the Frenkel–Kontorova model of crystal dislocations, we demonstrate the way to overcome the light speed barrier by introducing elvisebrions. Elvisebrions are defined as objects that can exist in the case when some hidden sectors, very weakly interacting with the visible sector, are either not Lorentz invariant, or Lorentz invariant but with different limiting velocities. In this case, elvisebrions can move in a superluminal manner without violating our existing physical models. 
\end{abstract}

\maketitle

\section{Introduction}
As 
one can 
understand, it is not an easy task to talk about breaking the light speed barrier, as it can raise various false rumors, especially after the sad story of the OPERA 
experiment \cite{1A}. 
However, after~having published a paper on the subject~\cite{1}, we decided to try to present it here in a rather entertaining~manner. 

\section{Experimental Basis of Special~Relativity}
Experimental confirmations of special relativity are ubiquitous~\cite{2}. One can say that special relativity is the most precisely tested theory. Let us provide some numbers to present the level of precision involved. Usually the breakdown of special relativity (Lorentz invariance) is characterized by introduction of Lorentz non-invariant terms in the interaction Lagrangian $L$. One such term 
is 
$\delta L\sim \vec{B}^{\,2}$. 
As~the separation of electromagnetic field into an electric $\vec{E}$ and a magnetic $\vec{B}$ fields depends on the observer, one should specify in which reference frame this square of the magnetic field is given. Usually the preferred frame is identified relative to the ``rest frame of the 
Universe'', 
$S$, in which cosmic microwave background is isotropic. In~this frame $S$, light velocity in vacuum, 
denoted by $c$, will 
differ from the maximum attainable speed of a material body, taken to be unity, due to the additional term in the 
interaction Lagrangian. In~another frame (for example, on~Earth), 
which moves with respect to the frame $S$ with velocity $\vec{u}$ (for the solar system, $u=371\pm 
1~\mathrm{km}~\mathrm{s}^{-1}$ in dimensionfull units
\cite{3}), 
 the~speed 
of light, $c$, 
 depends on the angle between its propagation direction and $\vec{u}$. This light propagation 
anisotropy in the laboratory frame should lead to observable effects that are proportional to $u^2(1-c^2)$ if Lorentz 
invariance is violated~\cite{4}.
 Nevertheless, no such anisotropy was found, and~these null results provide the limit of applicability for special 
relativity. The modern limit on the velocity difference is as small as $|1-c|<3\times 10^{-22}$ \cite{4}.

In addition to the square of the magnetic field, the~only other renormalizable term that affects the propagation of light is $\vec{A}\cdot\vec{B}\,s_0+(\vec{A}\times \vec{E})\cdot \vec{s}$ \cite{4}, where $s_\mu$ is a fixed 4-vector with a dimension of reciprocal length, 
the Greek letters take the values 0 (time) and 1, 2, and 3 (space coordinates), 
and~$\vec{A}$ is a vector potential. This term in Lagrangian violates not only Lorentz symmetry but also charge 
($C$),
 parity ($P$), 
 and~time ($T$) reversal 
symmetry ($CPT$). Astronomical data and particle physics experiments severely constrain the magnitude of components of $s_\mu$. Namely, the~constraint on the dimensionless parameter that might characterize $CPT$ violation in particle physics is $|s_\mu/m|<4\times 10^{-38}$ (with $m$ the electron mass) \cite{4}.

On the webpage~\cite{2} one can find not only numerous experiments that confirm special relativity, but~also several ones that contradict it. However, all experiments that refute special relativity have been wildly criticized, and~none of these experiments came anywhere close to making a convincing statement~\cite{2}. So, one can safely say that special relativity was thoroughly tested within the domain of its applicability and experimental evidence completely supports its~validity.

\section{Golden Shine of the Theory of~Relativity}
One could think that relativity belongs to the academic high-brow circles and has no relation to everyday life, but~it would be not more than an unfortunate fallacy. For~instance, every time 
one enjoys
glitter of gold, one meets 
special relativity!

The fact is that for heavy atoms, relativistic corrections are significant and even necessary to explain chemical and physical behavior of these atoms~\cite{5,6}. The~color of metals mainly results from the absorption of light when a $d$ electron ``jumps'' to an $s$ orbital. For~silver, the~absorption that corresponds to the $4d\to 5s$ transition requires energy of 3.7 eV which belongs to the ultraviolet range. As~a result, visible light is not absorbed by silver and all visible frequencies are reflected equally. This circumstance means that silver has no color of its own: it is silvery. If
one was 
to do corresponding calculations for gold without including relativistic effects, 
one 
would have predicted the same result: that gold should be silvery. However, due to relativistic effects the $s$ and $p$ atomic 
orbitals, 
in gold are more ``contracted'' compared to non-relativistic predictions (the 
so-called 
``relativistic stabilization''), while 
$d$ and $f$ orbitals are on the contrary destabilized and become more diffuse. Consequently, for~gold the gap between the $5d$ and $6s$ orbitals decreases and corresponds to the energy of 2.4 eV which leads to absorption of blue light and reflection of green and red. When combined, this spectrum results in the yellowish golden hue~\cite{6,7}.

One 
also meets 
relativity, both special and general, every time 
GPS (Global Positioning System) 
is used 
for navigation, because~without these theories GPS would not function properly~\cite{8}. Finally, even in 
rather 
old 
cars, 
special relativity
is under use! Indeed, calculations indicate that 10 out of 12 volts in a lead-acid car battery come from 
relativistic effects~\cite{9}.

\section{Why Are We Interested in the Light Speed Barrier?}
\begin{figure}[h]
\includegraphics[height=43mm]{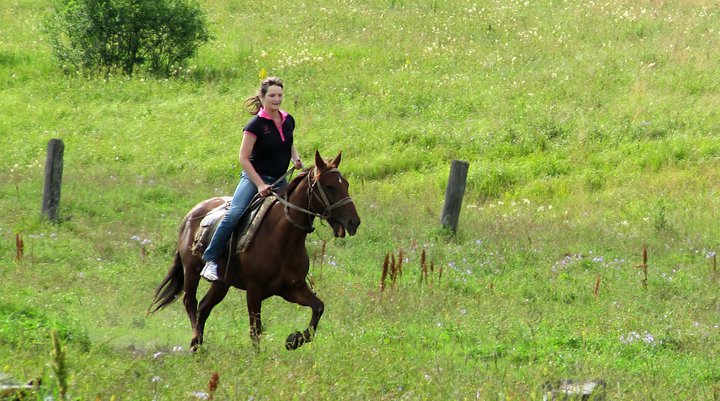}
\includegraphics[height=43mm]{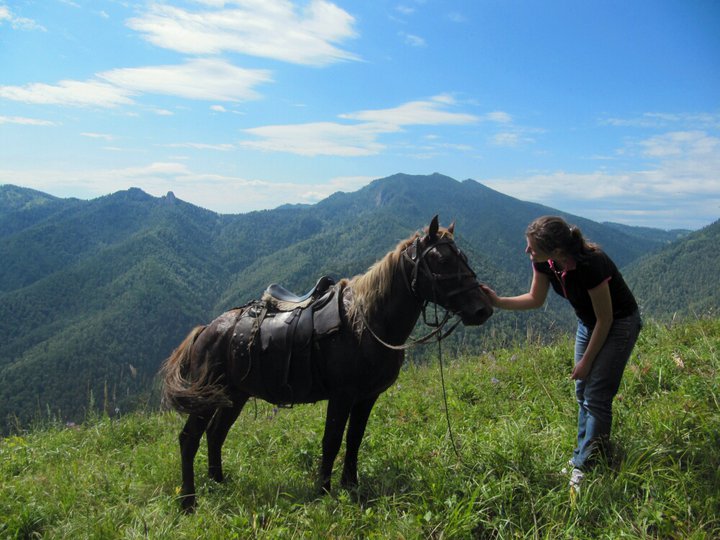}
\caption{Is it possible to break the light speed barrier on horseback?}
\label{horse}
\end{figure}

As 
one can
see, relativity reigns even in everyday life. Why then are we interested in questioning light speed barrier? After all, we 
have experimentally tested whether it is possible to overcome it. 
One of the authors 
gave her best try in a leisure holiday 
experiment, but~the results were overwhelmingly convincing: it is impossible to overcome light speed barrier on a horseback. 
 One 
can see that the horse was somewhat embarrassed by this negative~result 
(Figure~\ref{horse}). 

Quoting D.I. Blokhintsev~\cite{10}:
``The principal 
 reason for concentrating our attention specifically on the theory of 
relativity is, of~course, not any sympathy with ignorant attacks on this 
theory, but~the circumstance that its basic postulates touch upon the 
deepest foundations of the physics---geometry of space-time''.

There are psychological reasons too. A~human being cannot tolerate any limitations due to the rebellious nature of human 
mind. A~Russian writer, Vasily Shukshin, in~his short story 
{\it Stubborn fellow} 
describes this peculiarity of human mind very eloquently~\cite{11}: 
``Monya read the book which explained that a perpetuum mobile was 
impossible and that many had failed in an attempt to invent such a 
machine. He carefully studied the drawings of perpetual motion machines 
which had been proposed over the centuries. Then he started reflecting 
upon this problem, not troubling his head over friction or the laws of 
mechanics. He had his heart set on inventing a perpetuum mobile of a kind 
which had never been imagined before. For~some reason, he refused to 
believe that it was impossible.''

For the very same reasons, we also refused to believe that the speed of light barrier was really impossible to overcome. And~we set our hearts on inventing a way which had never been imagined before, a~way into a dreamland without the speed~limit.

\section{Why the Light Speed {Barrier?}}
\begin{figure}[h].
\includegraphics[width=0.5\textwidth]{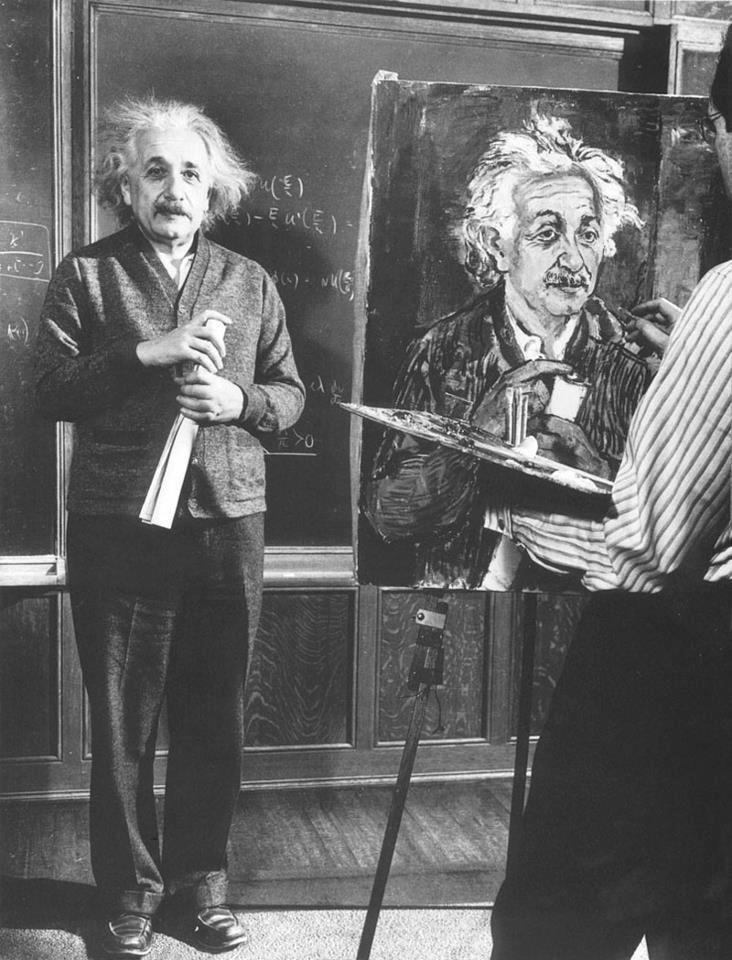}
\caption{Albert Einstein posing for a portrait painting.  
Taken from \cite{einstein_portrait}.} 
\label{einstein}
\end{figure}

{The 
} argument for the light speed barrier was given by 
Albert 
Einstein (Figure~\ref{einstein}) 
himself in his seminal first paper on special 
relativity ~\cite{12}.

This argument is 
transparent
enough: 
 relativistic expression for the kinetic energy is 
$$W=mc^2\left (\frac{1}{\sqrt{1-V^2/c^2}}-1\right ), $$ 
 with $m$ being the mass and $V$ being the velocity of the object. 

Then, following A. Einstein: ``Thus, 
 when $V=c$, $W$ becomes infinite. Velocities greater than that of light have, as~in our previous results, no possibility of 
 existence''~\cite{12}.

\section{For Light Spots, There Is No Light Speed~Barrier}
Not every kind of superluminal motion is excluded by Einstein's argument. One example of superluminal motion are light spots produced on a remote screen by a rotating source of light or particles. Such light spots are in many ways analogous to the subluminal radiation sources and in particular, they can generate Cerenkov radiation in "vacuum"~\cite{13}. Aspects of the superluminal motion were considered previously by Heaviside (1888), Sommerfeld (1904), Wigner (1939), Tangherlini (1958) and Tanaka (1960), to name a few~\cite{14}.

Major difference between the superluminal and subluminal radiation sources 
is that according to the Einstein's argument, the~first ones cannot be 
represented by individual particles, but~by patterns formed by aggregates 
of such particles in a coordinated motion. An~interesting example of 
practical application of superluminal motion is a Gyrocon---a high-power 
microwave generator in which radiation is produced by conically moving 
relativistic beam of electrons~\cite{14,15}. The~initial idea of a Gyrocon 
was suggested long ago independently by M.S. Neyman (1940), P.V. Hartley 
(1945) and J. W. McRae (1946) \cite{16}. However, the~first 
realizations~\cite{17,18} were imperfect and yielded very low power 
output. It was G.I. Budker who realized the possibilities offered by 
relativistic energies of the electron beam in such a device. In~a Gyrocon 
built by the Budker group, a~harmonic used for the generation had phase 
velocity which was 1.84 times larger than the light velocity. Output power 
of the Gyrocon in the continuous mode reached 5 MW and produced radiation 
in the wavelength range from 30 cm to 1.1 m with the efficiency of the 
beam energy conversion into radiation of about 80\% \cite{16}.

Superluminal motion of this type has long found industrial applications. However, this is not the superluminal motion that interests us~here.

\section{Does Somebody Live to the North of the Himalayas?}

In fact, Einstein's argument given above is misleading. 
Following  E.C.G. Sudarshan~\cite{19}: ``As analogy we 
may consider demographers studying the population patterns of the Indian 
Subcontinent. They would find that there are a lot of people in India but 
as you approached the Himalayas there are fewer and fewer people. Climbing 
the giddying heights is a difficult task and it is well-nigh impossible to 
climb it let alone live there. Suppose a demographer calmly asserts that 
there are no people north of the Himalayas since none could climb over the 
mountain ranges. That would be an absurd conclusion. People of Central 
Asia are born there and live there: they did not have to be born in India 
and cross the mountain ranges.''

\section{Tachyons}
 In 1962 Sudarshan, Bilaniuk and Deshpande introduced a notion of 
faster-than-light particles that live ``to the north of the Himalayas'' in 
the sense that they always move with the velocity greater than the 
velocity of light~\cite{20}. Later, such particles were named tachyons by 
Gerald Feinberg~\cite{21}.

One may wonder why it took so much time for the tachyon hypothesis to emerge. The~answer lies partly in the sociopsychology of science. ``In general, mathematicians tend to behave like "fermions" i.e.,~avoid working in areas which are too trendy whereas physicists behave a lot more like "bosons" which coalesce in large packs and are often "overselling" their doings, an~attitude which mathematicians despise''~\cite{22}. What is particularly worrisome is that ``this "bosonisation" of crowds of theoretical physicists does not make them tolerant to new ideas that don't fit with the common hopes of the "believers"''~\cite{23}.

In reality, the~tachyon hypothesis can be traced back much earlier, to works by Lev Yakovlevich Shtrum in 
1923~\cite{24,25}. Shtrum 
was a head 
of 
 Theoretical Physics Department 
at Kiev University from 1932 to 1936. Then, he was arrested and executed. His love for philosophy played a tragic role in his story, as it put him in touch with the philosopher Semyon Semkovsky, Lev Trotsky's cousin. Shortly after Semkovsky was arrested as Trotsky's relative, Shtrum was arrested too. He was tortured, forced to sign self-incriminations, and executed with a group of other Ukrainian scientists. The~Stalinist government tried to annihilate him spiritually as well: his works were removed from Soviet libraries and destroyed. As~a result, the~tachyon hypothesis fell into oblivion for many~years.

Risking his life, his friend and contemporary Vasily Grossman resurrected Shtrum's name in a novel that later became the first part of his renowned novel {\it Life and Fate}. The~novel was first published in 1952 when Stalin was still alive. Fortunately, nobody among censors had suspected that the main hero of the novel, Victor Shtrum, had a real prototype in the face of the executed physicist. Nevertheless, on~the grounds of official antisemitism of that time, censors demanded to remove the character from the novel. Grossman was steadfast with his answer: ``I agree with everything but Shtrum''~\cite{26}.

Interestingly, Nathan Rosen, one of Einstein's co-authors, replaced Shtrum as 
head of 
 Theoretical Physics Department
for two years. Rosen did not 
speak 
Russian and his lectures for students were translated from English. While working in the USSR, Rosen nevertheless learned to read a little Russian and fortunately, understood that he needed to flee the country when the events of 1937~began. 

When  
an unemployed American scientist Nathan Rosen found work in the 
USSR, 
he was most apparently 
 unaware of the tragic fate of his predecessor. 
In~fact, Rosen was 
a very modest and positive person, as~the following memories witness. 
A. 
Gordon recalls:~\cite{27}:
``I walked along the corridor seeing everywhere signs "Professor ...". 
At~the very end of the corridor I noticed a door with a small copper plate with a name without the professor title. The~office door opened and out of it came an old thin man politely greeting me. I saw him for the first time. Obviously, he was not a professor, maybe he was a manager, an~engineer, or~a technician. Professors, if~they greeted you, they greeted conscious of their high position in the world of science, so that you feel their significance and importance. The~owner of this office was too simple and friendly to be a professor. When he retired from the office, I approached the door to read the name on a small nameplate. It was written in a foreign language: N. Rosen.''

\section{Tolman's Antitelephone~Paradox}
\begin{figure}[h]
\includegraphics[width=0.6\textwidth]{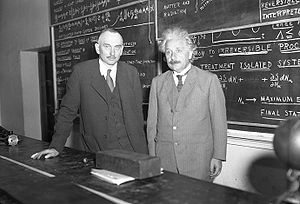}
\caption{Richard Tolman with Einstein at California Institute of Technology in 1932. 
Taken from \cite{tolman-einstein}.} 
\label{Tolman}
\end{figure}

The interval between emission and absorption of a superluminal tachyon is spacelike and this circumstance creates a real problem for the tachyon hypothesis: for events separated by a spacelike interval, their relative time order depends on choice of a reference frame, and~in some inertial reference frames the tachyon will be absorbed before it is emitted. In~1917, Richard Tolman (shown with Einstein in 
Figure~\ref{Tolman})
in his book {\it The Theory of Relativity of Motion} showed that in this case 
one faces 
a paradox of causality, today 
called Tolman antitelephone paradox. Figuratively speaking, one can send a message into one's own past using~tachyons.

In fact, Tolman's antitelephone paradox has already been invented in 
1907 by Einstein himself~\cite{28}, who looks was 
aware of the 
weakness in the infinite kinetic energy argument. He 
concludes~\cite{28}: 
``This 
result signifies that we would have to consider as possible a transfer 
mechanism whose use would produce an effect which precedes the cause 
(accompanied by an act of will, for~example). Even though, in~my opinion, 
this result does not contain a contradiction from a purely logical point 
of view, it conflicts so absolutely with the character of all our 
experience, that the impossibility of the assumption $V>c$ is sufficiently 
proven by this result''.

The issue, raised by Tolman's antitelephone paradox is, however, already 
present in quantum field theory that unifies the fundamental ideas of 
special relativity and quantum mechanics and is at the base of our modern 
theories of elementary particle~physics.

\section{Can Quantum Theory and Special Relativity Peacefully Coexist?}
\begin{figure}[h]
\includegraphics[width=0.7\textwidth]{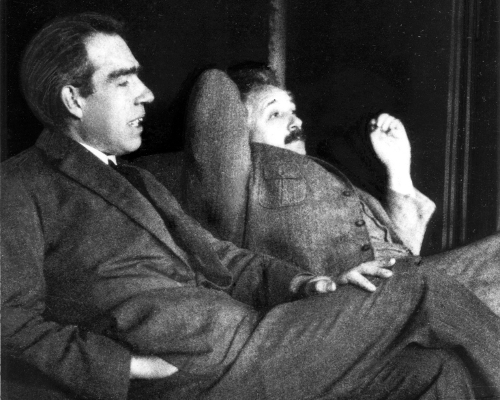}
\caption{Niels Bohr and Albert Einstein in 1925. Taken from \cite{bohr-einstein}.}
\label{bohr_einstein}
\end{figure}

The second and most profound revolution in physics after special 
relativity was creation of quantum mechanics 
(Figure~\ref{bohr_einstein} 
shows Bohr and Einstein discussing the point).  
Both theories are very successful. However, they are based on entirely different ideas, which are not easy to reconcile with each~other.

Following F. Wilczek \cite{29}:
 ``Imagine a particle moving on average at nearly 
the speed of light, 
but~with an uncertainty in position, as~required by quantum theory. Evidently, there will be some probability for observing this particle to move a little faster than average, and~therefore, faster than light, which special relativity will not permit.'' 

\section{Quantum Theory and the Problem of Superluminal~Motion}
In quantum field theory, the~amplitude for a particle to propagate from a space-time point $x$ to a point $y$ is Lorentz invariant and is given by the Wightman propagator. When the 
difference $x-y=(0,\vec{r})$ is purely in the spatial direction, 
and~when the distance 
 $r$  
is much smaller than the particle's reduced Compton 
wavelength 
$m^{-1}$ 
(natural units $\hbar=c=1$ 
are 
assumed,
 where $\hbar$ is the reduced Planck constant), 
the~Wightman function has the following asymptotics, 
 $D(x-y)\approx \frac{m}{4\pi^2 r} \,\sqrt{\frac{\pi} {2mr}}\,e^{-mr}$, 
and is not zero, despite zero 
time-interval between events at 
coordinates 
$x$ and $y$. This means that the particle has a non-zero probability to 
propagate with infinite velocity in this frame. As~
one can  
see, it is not a trivial task to reconcile quantum mechanics with its 
notorious non-localities with special relativity. 
Quoting J. Butterfield~\cite{30}: 
``Relativistic causality---formulate it as you like!---is a subtle matter 
in relativistic quantum theories.''
Let us mention some other examples of the alleged superluminal effects 
which demonstrate a subtlety of coexistence of quantum mechanics with special~relativity.

Hegerfeldt's theorem states that the wave function of a quantum system, if~initially localized in a finite region, immediately develops an infinite tail spreading over all space. As~a result, transition probabilities in widely separated systems may instantaneously become nonzero~\cite{31}. Maybe, then, it is not surprising that entanglement and mutual correlations can also be generated at space-like separated points~\cite{32}. All of these are, of~course, in~spirit of old Einstein, Podolsky and Rosen paradox~\cite{33}.

However, it becomes even worse, as quantum mechanics predict that the 
transmission time across a potential barrier becomes independent of 
barrier thickness for very thick barriers, the~so-called 
``Hartman effect''~\cite{34}. 
Let us also mention the Scharnhorst and 
Drummond--Hathrell 
effects. The~Scharnhorst effect  
predicts that light signals travel a bit faster than $c$ 
in the Casimir vacuum between two closely spaced conducting plates~\cite{35}. In~the 
Drummond--Hathrell 
effect, photons may propagate with superluminal velocity in certain gravitational backgrounds~\cite{36,37}. 
It is believed,  
however, that neither of these strange effects allows 
information to be transmitted faster than the light velocity, and thus, they are compatible with special relativity, although in a very subtle~way \cite{37A,37B}.

\section{Frustrated Total Internal~Reflection}
Frustrated total internal reflection~\cite{38} is an optical analog of the quantum mechanical tunneling. It is 
known 
that, if an angle of incidence of a light beam at the boundary that 
separates optically denser medium, in~which the light beam propagates, and~the one of lower density, exceeds some critical angle, then it is totally reflected at this boundary. However, even in this case the electric field penetrates the adjacent medium, with~the penetration depth of order of a wavelength, in~a form of an evanescent wave. Evanescent modes are classical analogs of virtual photons~\cite{39}. If~a third optically dense medium is added at around a wavelength from the first medium, then because of the evanescent wave, the~light beam penetrates to this medium much like the quantum tunneling of particles through potential~barriers.
\begin{figure}[h]
\includegraphics[width=0.5\textwidth]{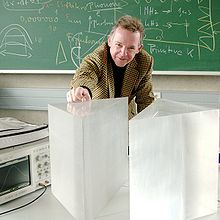}
\caption{G\"unter Nimtz in the Physics laboratory of the University of Koblenz, 2008. 
Taken from 
\cite{nimtz-pic}.
} 
\label{Nimtz}
\end{figure}

The Hartman effect was experimentally confirmed in frustrated total 
internal reflection~\cite{40}. G\"unter 
Nimtz 
(Figure~\ref{Nimtz}) 
even states that he and his collaborators managed to transfer modulated Mozart's 40th symphony at a superluminal speed of 
 4.7$c$ 
 using this phenomenon for microwaves~\cite{41}. Nevertheless, it seems that he is at error~\cite{42,43}. If~you define the signal velocity as the velocity of the pulse maxima then indeed, an~evanescent barrier can be crossed superluminally. However, even these superluminal wave packets still obey causality and the peaks of pulses in front and behind the evanescent barrier are not causally related~\cite{44}.

\section{There Are No Paradoxes in~Physics}
\begin{figure}[h]
\includegraphics[width=0.7\textwidth]{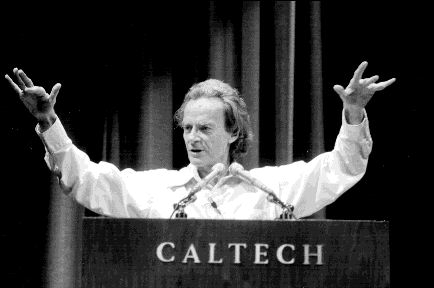}
\caption{Richard Feynman lecturing at Caltech Seminar Day, 1978. Taken from \cite{feynman-pic}.} 
\label{feynman}
\end{figure}

``Of course, in~physics there are never any real paradoxes because there 
is only one correct answer; at least we believe that nature will act in 
only one way (and that is the right way, naturally). So in physics a 
paradox is only a confusion in our own understanding,'', quoting 
 R. Feynman
(Figure~\ref{feynman})~\cite{45}. 
\unskip

\section{The Idea of~Antiparticles}
The above-given  
examples illustrate that there is some tension 
between 
quantum theory and special relativity in what concerns superluminality. 
Quoting F. Wiczek~\cite{29}: 
``The only known way to resolve this tension involves introducing the idea 
of antiparticles.''

Quantum field theory offers a miraculously clever solution to the 
superluminal propagation dilemma \cite{45A}. 
It is true that, for a particle propagating superluminally between space-time points separated by a spacelike interval, there is always a reference frame in which the particle propagates backwards in time, so that its absorption preceding its emission is in apparent violation of causality. Nevertheless, the~particle's energy in this frame is negative, and~a negative energy particle propagating backward in time can be interpreted as a positive energy antiparticle propagating forward in time. This Stueckelberg-Feynman interpretation of antiparticles as negative energy particles moving backward in time provides quantum field theory's resolution of superluminal propagation dilemma. Causality demands that any measurement performed at a space-time point $x$ cannot influence an outcome of another measurement at space-time point $y$ if the points are separated by a spacelike interval. Mathematically this means that two local observables commute if separated by a spacelike interval. The~existence of antiparticles is crucial for a cancellation of all acausal terms in commutators of two local observables at spacelike separation. Therefore, quantum field theory does not allow information to be transmitted faster than the speed of light but at the expense of introducing~antiparticles.

\section{Stueckelberg---An Unconventional~Hero}
There was one central figure in the development of the idea of antiparticles as negative energy particles moving backwards in time. His full name was Johann Melchior Ernst Karl Gerlach Stueckelberg, Freiherr von Breidenbach zu Breidenstein und~Melsbach. 

\begin{figure}[h]
\includegraphics[height=60mm]{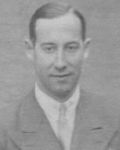}
\caption{Ernst Carl Gerlach Stueckelberg in 1934 at London International Conference on Physics. 
Taken
from \cite{stueckelberg34}.} 
\label{stueckelberg}
\end{figure}

Stueckelberg 
(Figure~\ref{stueckelberg})
 was from an aristocratic Swiss family. He made some deep and fundamental 
discoveries but remained a widely under-appreciated genius in the history of science~\cite{46}. Maybe this is explained by the fact that apart from his relationship with Wolfgang Pauli, Stueckelberg had little contact with the scientific community partly due to his mental illness~\cite{47}. While being in Princeton in 1932, Stueckelberg suffered from the first attack of bipolar disorder. Later during his life, the~recurrent psychiatric conditions resulted in countless stays in mental hospitals and seriously handicapped Stueckelberg's scientific career~\cite{47}.

Stueckelberg was not only the first person who introduced the 
aforementioned interpretation of antiparticles and illustrated the concept 
with graphs of space-time trajectories which 
today 
are called
``Feynman 
diagrams'', he was also years ahead in the development of quantum 
electrodynamics (QED). Unfortunately, he did not promote his ideas, and 
published his articles mostly in French
in a Swiss journal {\it Helevetica Physics Acta}
 with a 
small international~audience \cite{50}.

After receiving the Nobel Prize 
for his work on QED, 
Richard Feynman gave a lecture in Geneva at which Stueckelberg was 
present. After~the lecture, Feynman surrounded by his admirers noticed 
that Stueckelberg quietly left the lecture hall accompanied by his dog. 
Feynman was moved and remarked~\cite{48, 49}:
 ``He did the work and walks alone towards the sunset, and~here I am, 
covered in all the glory, which rightfully should be his.''

Stueckelberg's dog, Carlo III, was as legendary, as~his master. Stueckelberg attended seminars and lectures always accompanied by his dog. Carlo III had even been granted right of access to CERN by the 
CERN Director General~\cite{46}.  
At~seminars,
 Carlo III had a reputation of a great expert in theoretical 
physics. Usually the dog was well behaved watching the lecture while 
Stueckelberg napped. Occasionally. Carlo III barked awaking his master who then always spotted a mistake on the blackboard~\cite{46,50}. The~only reasonable explanation of how the dog spotted the errors (suggested by Valentine Telegdi~\cite{46}) is that Carlo III was not an expert in theoretical physics but was a great expert in his master's mood and he was reacting to his master unconscious~uneasiness.

\section{Reinterpretation~Principle}
Let us return to tachyons. One may think that the reinterpretation 
principle solves Tolman antitelephone and similar paradoxes~\cite{51,52}.

\begin{figure}[h]
\includegraphics[height=60mm]{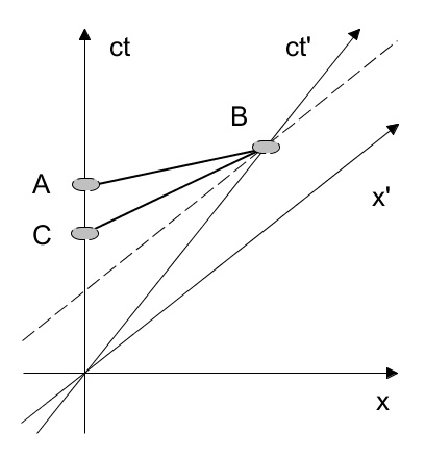}
\caption{Antitelephone paradox (see 
text for details).} 
\label{antitelephone1}
\end{figure}

Let two observers $\tilde A$ 
and $\tilde B$
 move with constant subluminal relative velocities, 
$t$ and $t^\prime$ axes being the corresponding world-lines,
 see Figure~\ref{antitelephone1}. Observer 
$\tilde A$ 
sends a tachyon that follows $AB$ world-line (moves forward in time with 
respect to $\tilde A$) and is immediately reflected upon arrival at point $B$. 
World-line $BC$ lies above the dashed line (the line of simultaneity for 
the event $B$) which is parallel to the coordinate axis $x^\prime$. 
Therefore, with~respect to 
$\tilde B$, 
the~reflected tachyon moves forward in time. However, with~respect to 
$\tilde A$,
it moves backwards in time and arrives at point $C$ earlier than it was emitted. 

The resolution of this paradox according to the reinterpretation principle is the following. It can be shown that while moving from $B$ to $C$, the~tachyon has a negative energy. Therefore, it can be reinterpreted as an antitachyon with a positive energy moving forward in time. So, from~the point of view of the observer $A$ the sequence of events looks like this: an antitachyon is emitted at point $C$ and then a tachyon at point $A$. Both have positive energies and move forward in time. Then they annihilate at point $B$. There is thus no paradox in this sequence of~events.

Not 
everyone 
agrees with such a solution to the Tolman's 
antitelephone paradox (see, 
e.g., 
~\cite{53}). 
Quoting Ref.~\cite{54}:
``Causality objections against superluminal particles are by far the most 
subtle, and~much room for reflection remains in this regard.''

\section{Gell-Mann's Totalitarian~Principle}
Leaving aside subtleties related to the causality, 
 one  
can invoke Gell-Mann's totalitarian principle 
``everything 
not forbidden is compulsory'' and ask: If 
tachyons are not forbidden by special relativity, why are they not observed? (In fact, the~totalitarian principle first appeared in T.H. White's Arthurian fantasy novel 
{\it The Once and Future King}  
~\cite{55}). Up~to here, 
superluminality 
was emphasized 
a defining property of tachyons. This is no longer justified in quantum theory. Classical concept of propagation velocity is not well-defined for evanescent modes or virtual photons, as~nothing well defined and localized propagates through the tunneling barrier in these~cases.

\begin{figure}[h]
\includegraphics[height=20mm]{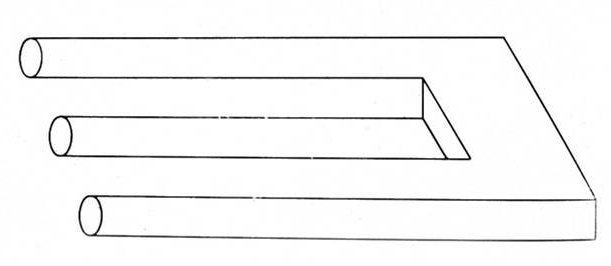}
\caption{Devil's pitchfork -- a classic impossible figure.} 
\label{devilsPitchfork}
\end{figure}

The notion of particles, borrowed from the classical physics, is also not quite adequate in the quantum world. As~a result, 
 one can still speak about wave-particle duality, a~concept as 
dubious 
as the devil's pitchfork: a classic impossible figure 
(Figure~\ref{devilsPitchfork})~\cite{56}.

Left and right parts of this figure may be interpreted as two-dimensional projections of three-dimensional bodies, much like a part of properties of a quantum particle may be interpreted as particle-like while another part as wave-like. Then 
one tries 
to unify these interpretations through the concept of wave-particle duality, pretending that the figure above is a true three-dimensional figure. However, such interpretation is impossible, as the figure is, in reality, in two dimensions~\cite{57}.

So, it is better to call a quantum particle a quanton, which is neither a 
particle nor a waive, but a completely new object unknown to classical 
physics~\cite{58}. The~old tradition of describing the quantum world in 
classical concepts creates, of~course, a~serious logical tension when 
using two contradictory ideas for describing the same object, even if the 
so-called complementarity principle is evoked to soften conceptual 
conflicts. ``It must be realized today that this view of the quantum 
world, adapted as it was to its first explorations, is totally outdated. 
In~the past fifty years, we have accumulated sufficient familiarity, 
theoretical as well as experimental, with~the quantum world to no longer 
look at it through classical glasses,'' quoting  
J.-M. L\'{e}vy-Leblond~\cite{58}.

\section{Classification of~Quantons}
The best-known way to classify free quantons propagating in Minkowski spacetime is through their symmetry properties, assuming that elementary quantons correspond to the unitary irreducible representations of the 
Poincar\'{e} group (inhomogeneous Lorentz group) found by Wigner. Depending on the four-momentum $p$ of a quanton, whose square is Lorentz invariant, and~on the sign of its energy, 
one has 
six different classes of such representations~
\cite{59}:
\begin{enumerate}
\item $p^2>0$, $p^0>0$,
\item $p^2>0$, $p^0<0$,
\item $p^2=0$, $p^0>0$,
\item $p^2=0$, $p^0<0$,
\item $p^\mu=0$,
\item $p^2<0$.
\end{enumerate}

Representations with negative energies (classes 2 and 4),
as~was already mentioned, 
 should be reinterpreted as corresponding to antiparticles with positive 
energies belonging to 
classes 1 and 3.
Class 1 
corresponds to bradyons, massive quantons, which always move with 
velocities less than the speed of 
 light 
 in vacuum if described from classical 
 perspective. 
 Class 3 
 contains representations corresponding to 
luxons, quantons with zero mass like photons, which always move with 
maximum velocity $c$.

Representations in each group are further classified by considering the 
unitary irreducible representations of the so-called little group, 
a~subgroup of the Poincar\'{e} group that leaves invariant a particular 
four-momentum characteristic for a given class. For
class 1, 
 a rest-frame momentum $p^\mu=(m,0,0,0)$ 
 can be chosen,  and the little group is 
the three-dimensional rotation group SO(3). 
This group  
has finite dimensional 
unitary representations, characterized by one real number---the spin of 
the 
quanton. For~
class 3, 
$p^\mu=(k,0,0,k)$ can be chosen, and the 
little group is the Euclidean group ISO(2) or transformations of the 
Euclidean plane which preserve the distance between points. ISO(2) is a 
semi-direct product of SO(2) and T(2), a~group of translations in the 
plane. Correspondingly, the~unitary irreducible 
representations, 
corresponding to the massless quantons, are of two types. If~T(2) is 
represented by trivial representation, then unitary irreducible 
representations correspond to the representations of the two-dimensional 
rotation group SO(2) and are finite dimensional and characterized by the 
helicity of the quanton. If~T(2) is represented by non-trivial 
representation, then the corresponding unitary irreducible representations 
of the Poincar\'e group are infinite dimensional and are characterized by 
a continuous parameter with the dimension of a mass even if they describe 
massless quantons. The~second exotic property of such representations is 
that the corresponding quantons have infinite degrees of freedom (these 
representations contain all helicities from $-\infty$ to $\infty$ and are 
called infinite spin representations). Wigner has 
argued that the heat 
capacity of a gas of such quantons would be also infinite and rejected 
such representations as unphysical. For~the modern perspective on the 
infinite spin representations, see, e.g,
~\cite{60}.

Class 5 
are representations of the symmetries 
of the vacuum. In~this case, the little group is the entire Lorentz group, 
and all unitary representations are infinitely dimensional. 
 These represantations 
 are 
organized into the main and supplementary series. 
Some of them 
can correspond 
to pomerons, ``without a doubt, one of the weirdest concepts 
of something particle-like that one could imagine,'' quoting 
J. Swain~\cite{59}.

Finally, 
 class 6 corresponds to tachyons. The~little group 
is 
 SO(2,1), the~Lorentz group in a spacetime with one time-like and two 
spacelike dimensions. It is non-compact and has only infinite dimensional 
unitary representations except for a one-dimensional trivial 
representation. Thus, tachyons are quite different from the usual bradyons 
and luxons: either tachyon is a scalar particle with no spin, or~it has an 
infinite number of polarization states. However, it was argued that, 
for 
tachyons, 
 it 
is justified to resort to nonunitary representations, which 
are finite dimensional. 
 This  
 way integer and half-integer spins can be 
associated to tachyons too~\cite{51} (although the finite-dimensional 
representations of the little group for tachyons are not unitary, a~fully 
covariant theory for tachyons with any spin can be constructed with a 
helicity dependent invariant integral which cannot have the usual 
interpretation of probability or charge~\cite{51A}).

\section{Tachyons and~Instability}
In quantum field theory, quantons are associated to relativistic fields \cite{51AA,51BB,51CC}. 
 For~the field $\phi$, the~squared mass of the corresponding 
quanton 
equals the second derivative of the self-interaction potential $V(\phi)$ of this field at the origin $\phi=0$. If~the squared mass is negative as for tachyons, the~origin cannot be the minimum of the potential. In~other words, the~system with a tachyonic degree of freedom is unstable at $\phi=0$ tachyonic vacuum state. As~a result, the~tachyonic field $\phi$ will roll down towards the true vacuum, acquiring a non-zero vacuum expectation value (tachyon condensation).

For the true vacuum, the~squared mass (the second derivative of the 
self-interaction potential) is positive, and~small excitations of the 
field $\phi$ around the true vacuum will appear as bradyons, 
not~tachyons (for a pedagogical discussion of mass generation through spontaneous symmetry breaking, see \cite{51B}). 

If 
one considers 
spacelike four-momentum as a defining property of tachyons, then it 
becomes clear that tachyons play an important role in modern physics. The~most notorious example of tachyon condensation is Higgs mechanism in the Standard Model of elementary particles physics. 
 One  
can say that the Higgs boson is the most famous would-be tachyon. Tachyon condensation is important in string theory too~\cite{61} and 
can  play 
a role in early cosmology~\cite{62}.

Finally, every quanton can become tachyonic in a virtual state (the four-momentum of a virtual quanton can be spacelike).

\section{Tachyons and the Light Speed~Barrier}
Initially, we were interested in tachyons as superluminal objects, but~a 
disappointment awaits us here: tachyons cannot support the true superluminal propagation, even in a rolling state towards true vacuum. The~argument goes as follows~\cite{62A,63,64}.

Propagation of a free scalar tachyon in Minkowski spacetime is described by the usual wave equation for spin-zero particles, the~Klein--Gordon equation, but~with imaginary mass. The~fundamental plane-wave solutions of the Klein--Gordon equation have the form 
$\phi(t,x)=\exp{(-iEt+ipx)}$, 
 where $E$ is the energy, 
for all relativistic quantons including tachyons for which $E^2-p^2=m^2$ 
with imaginary mass $m$. Because~of the unusual negative sign of the squared mass term, the~Klein-Gordon equation will have two kinds of solutions. For~the first kind, $|p|>|E|$ and $E$ is real. In~this case, 
 one has 
usual plane waves. For~the second kind, $|p|<|E|$,
 and then 
$E$ 
 becomes 
 imaginary. 
The~corresponding wave amplifies exponentially as time~passes.

If 
the second type solutions are allowed,
then 
a localized unstable wave-packet
can be built.
 However, the~Klein--Gordon equation, irrespective of the sign of the mass term, is a hyperbolic partial differential equation. Hyperbolic partial differential equations do not allow localized disturbances to spread with speeds faster than the maximum speed which is 
 here
 the speed of light. If 
the second-type solutions are excluded,
 then it  
will be 
no longer possible to build localized wave-packets but only non-local disturbances of the tachyonic field that cannot be used to send information from one place to another faster than the speed of~light.

Impossibility to construct a stable localized free-tachyon wave-packet implies that 
 one
cannot transmit information by tachyon field since the information would be destroyed by the inherent instability of these fields~\cite{63}. 
Quoting G.W. Gibbons \cite{62}:
``Contrary to popular prejudice: the tachyon is not a tachyon!'' 

\section{The Frenkel--Kontorova~Model}
It is time to change 
 the  
strategy in searching 
for superluminal objects. From~what was said so far, it should be clear 
that 
there is 
 virtually 
no chance if the Lorentz symmetry is indeed a fundamental property of 
Nature. On~the other hand, 
one can
consider situations where Lorentz symmetry emerges at low-energy limit but is not a fundamental property from the very beginning. 
The~Frenkel--Kontorova model of crystal dislocations provides a good 
starting 
point \cite{62A}. 
 The Frenkel--Kontorova model describes a one-dimensional 
chain of atoms with harmonic interatomic interactions subjected to an 
external sinusoidal 
substrate~potential. 

The remarkable fact about the Frenkel--Kontorova model is that in the long 
wavelength approximation, when the lengths characterizing the chain 
discreteness are much smaller than the wavelength of chain excitation, its 
dynamics are described by the so-called sine-Gordon equation,
$$\frac{1}{c^2}\frac{\partial^2 \Phi}{\partial t^2}-\frac{\partial^2 \Phi}{\partial x^2}+m^2\sin{\Phi}=0,$$
where $\Phi$ is the dimensionless field of displacements of the individual atoms from their equilibrium positions (see \cite{1} for details). 

If the external potential is switched off, $m=0$ and the equation 
describes massless phonons traveling at the sound velocity, $c_s=c$. 
For~small 
oscillations $\Phi\ll 1$ and with non-zero external potential,
 the 
sine-Gordon equation turns into the Klein--Gordon equation 
describing 
massive phonons (bradyons) moving with subsonic~velocities.

One 
can also consider small oscillations around the point of unstable equilibrium $\Phi=\pi$. 
Setting 
$\Phi=\pi-\varphi$ and assuming $\varphi\ll 1$, 
 one obtains
the equation  
$$\frac{1}{c^2}\frac{\partial^2 \varphi}{\partial t^2}-\frac{\partial^2 
\varphi}{\partial x^2}-m^2\,\varphi=0,$$
which has a minus sign before the mass term and describes supersonic 
phonons (tachyons). Despite the supersonic behavior, the~field $\varphi$ does not allow information to be transmitted with the velocity $V>c$, as~was explained above. 
 One 
can repeat the basic arguments using this 
very 
example. The~dispersion relation linking the frequency $\omega$ to the wavenumber $k$ of tachyonic excitation has the form $\omega=c\sqrt{k^2-m^2}$. When $k<m$, $\omega$ becomes imaginary, indicating the onset of instability. Any sharply localized source of perturbation will have such wave numbers in its Fourier spectrum and cause the instability. It can be then shown that the localized disturbance (information) will propagate with velocity  $V<c$ in accordance with the Cauchy--Kowalewska theorem of hyperbolic partial differential equations~\cite{65}.

\section{Frenkel--Kontorova~Solitons}
\begin{figure}[h]
\includegraphics[height=27mm]{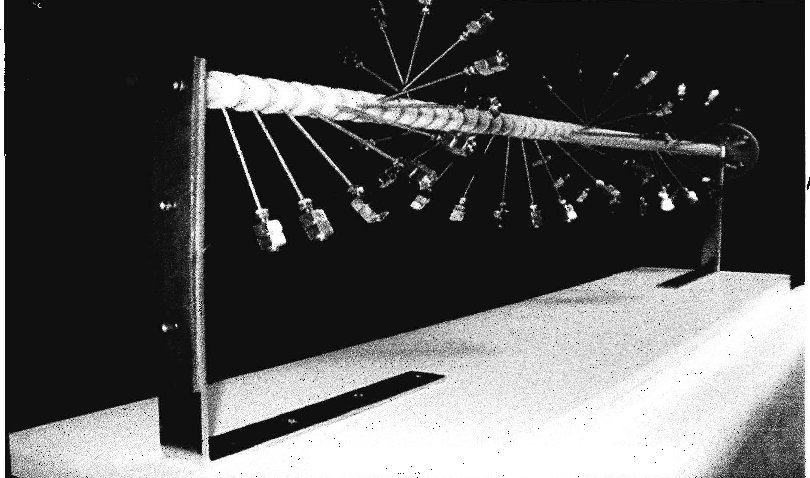}
\caption{Chain of pendulums as a mechanical analogue of the Frenkel--Kontorova model.} 
\label{pendulums}
\end{figure}

The Frenkel--Kontorova model has a mechanical analog: a chain of pendula 
which can rotate around an elastic horizontal axis that couples them to 
each other by exerting a harmonic~torque, see Figure~\ref{pendulums}. 

Some excitations of this chain of pendula have a form of traveling stable waves called solitons. Two particular cases of solitons are a 
kink ($\sigma=1$) 
and anti-kink ($\sigma=-1$): 
$$\Phi(x,t)=4\arctan{\left (\exp{\left [\frac{\sigma}{\lambda}\,\frac{x-Vt}{\sqrt{1-V^2/c^2}}\right ]}\right )}.$$

\begin{figure}[h]
\includegraphics[width=0.8\textwidth]{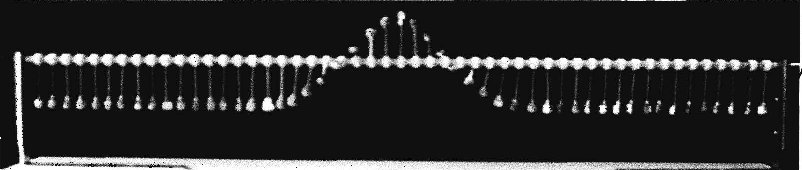}
\caption{Mechanical model of the Frenkel--Kontorova kink.} 
\label{kink}
\end{figure}

Here, $\sigma=\pm 1$, $V$ is a kink velocity and $\lambda$ is a Compton wavelength 
(m$^{-1}$) 
of the Frenkel--Kontorova phonons. As~one can 
see from 
the above 
expression, the~width of the kink depends on its velocity the same way as 
for relativistic particles. Therefore, it always moves with 
subsonic~velocities; see 
Figure~\ref{kink} for a mechanical model of the Frenkel--Kontorova kink.

On the other hand, 
there is  
a similar solution of the sine-Gordon equation, which can be called T-kink, and which always moves with supersonic velocity:
$$\Phi(x,t)=\pi-4\arctan{\left (\exp{\left [-\frac{\sigma}{\lambda}\,\frac{x-Vt}{\sqrt{V^2/c^2-1}}\right ]}\right )}.$$

It is evident 
that T-kink is a mechanical model of a tachyon. In~contrast to kinks, T-kinks are not stable. 
The 
reason for this difference can be understood 
as follows~\cite{1}. 
One 
can 
visualize infinitely many minimums of the Frenkel-Kontorova external potential as depressions of a long slate. A~kink corresponds to an infinite rope which begins in one depression of the slate and ends up in another neighboring depression. For~an infinite rope, one needs an infinite energy to throw the rope from one depression to another one. For~this reason, the~kink is stable (can be annihilated only by a counter anti-kink). On~the contrary, T-kink corresponds to a rope which lays on a slate ridge, then somewhere on the ridge it falls in the depression and raises again to the adjacent ridge. Of~course, such configuration cannot be~stable.

\section{Emergent~Relativity}
\begin{figure}[h]
\includegraphics[height=30mm]{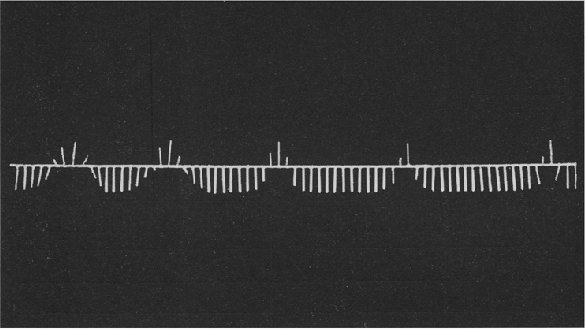}
\caption{Visualization of the Lorentzian contraction of the width of the Frenkel--Kontorova kink (see 
 the 
 text for 
 explanation).} 
\label{lengthcontraction}
\end{figure}

Frenkel--Kontorova solitons, remarkably, exhibit relativistic behavior. 
 It was altready seen 
 that the kink's width is subject to Lorentz contraction. Interestingly, 
such length contraction can be observed by naked eyes on the strobe 
photography of a pendula realization of a kink traveling with some 
dissipation of energy due to friction effects~\cite{66}, see 
Figure~\ref{lengthcontraction}.

The kink moves from right to left and becomes wider as its velocity 
decreases. One can calculate the kink energy and discover that it is also 
given 
 by the 
 known relativistic expression, 
 $E={Mc^2}/{\sqrt{1-V^2/c^2}}$, 
where $M$ is some constant (the kink 
mass) determined by parameters of the Frenkel--Kontorova model~\cite{1}.

This emergence of relativistic expressions is not surprising, as the sine-Gordon equation is in fact Lorentz invariant, provided the light velocity is replaced by the 
sound velocity $c$. What is truly remarkable is that this relativistic invariance is an emergent phenomenon. 
It is absent at the fundamental level of the Frenkel--Kontorova chain, but appears in the long-wavelength limit. Apart from the Frenkel--Kontorova model, there are other examples when a relativistic behavior emerges in purely classical mechanical systems~\cite{67}.

Mexican artist Octavio Ocampo's painting {\it Mouth of flower} 
\cite{flower}
illustrates the idea of emergence wonderfully well. 
 One can
see  
the face of a lady on the painting. However, at~a~closer inspection 
one finds 
that at the fundamental level, 
 there is 
no face but only some flowers and a~butterfly. 


\section{Supersonic~Solitons}
Emergent relativity in the Frenkel--Kontorova model is approximate and 
holds only insofar as  
 one  
can neglect discreteness effects. Even in the long-wavelength limit, not 
all interatomic interactions lead to the emergence of sound speed 
relativity in the atomic chains of the Frenkel--Kontorova type. Morikazu Toda 
managed to find exponential interatomic interactions that support supersonic~solitons.

When the velocity of a Toda soliton approaches the sound velocity, its behavior is opposite to that of a tachyon: its energy approaches to zero and its width approaches to infinity. Despite this difference, the~final result is the same: Toda solitons, like tachyons, cannot cross the sound barrier and become~subsonic.

If 
one 
generalizes 
the Frenkel--Kontorova model by adding special kinds of 
anharmonicity in the interatomic potential, other types of solitons can be 
constructed that can cross sound barrier \cite{67}. In~contrast to 
Frenkel--Kontorova and Toda solitons, Kosevich--Kovalev solitons can move 
with any velocity from zero to infinity~\cite{68}.
 In~some sense they 
interpolate between the subsonic Frenkel--Kontorova solitons and supersonic Toda solitons: for velocities much smaller than the sound velocity they resemble Frenkel--Kontorova solitons, and~for velocities much greater than the sound velocity they resemble Toda~solitons. 

\section{Elvisebrions}
 The above 
analyis of Kosevich--Kovalev and Toda solitons suggests a way how superluminal objects may be introduced in our world. Let us explain by a vivid analogy. 
 One
can classify creatures according to where they live with regard to the Himalayan range (light speed barrier).

 There are 
creatures that live south to the Himalayas (bradyons), that hover over the range (luxons), and~that live to the north (tachyons but they tend to condensate and move to southern regions). However, 
 one 
cannot exclude that one day
 one meets 
an alien (though 
 never  
met any of them 
so far)  
from a planet with no Himalayan range at all!

Such type of possibly superluminal objects, analogs of the Kosevich--Kovalev and Toda solitons, is conceptually different from the tachyons and deserves its own name. We name such objects elvisebrions (elvisebri 
means 
``swift as a lightning flash'' in Georgian). 

In more formal wording, 
we suggest a possibility that a hidden sector may exist which is either not Lorentz invariant or it is Lorentz invariant but with a different limiting speed. If~the two sectors (hidden and visible) are connected very weakly, then it is expected that the Lorentz invariance may still be a very good approximation in the visible sector. Nevertheless, such a setup is no longer Lorentz invariant as a whole and, therefore, a~possibility of hidden sector-induced superluminal phenomena does~appear.

The two sectors could be connected through the Higgs portal as considered 
in~\cite{69} and further analyzed in~\cite{70}. It was shown that Lorentz 
invariance violation in the Higgs sector is communicated at one loop level 
to all other Standard Model particles. However, in~the framework of the 
model, such effects are naturally small. Similar ideas have been 
formulated by Gonzalez--Mestres who calls this type of elvisebrions 
superbradyons~\cite{71}.

Finally, let us point to an 
interesting 
 paper 
by Robert Geroch~\cite{72} which actually provides a much more solid foundation for the elvisebrions hypothesis than we were able to give in this informal~essay. 

Quoting \cite{72}:
``In short, the~causal cones of special relativity, from~this perspective, have no special place over and above the cones of any other system. This is democracy of causal cones with a vengeance. This, of~course, is not the traditional view. That view---that the special relativity causal cones have a preferred role in physics---arises, I suspect, from~the fact that a number of other systems---electromagnetism, the~spin-s fields, etc.---employ precisely those same cones as their own. And, indeed, it may be the case that the physical world is organized around such a commonality of cones. On~the other hand, it is entirely possible that there exist any number of other systems---not yet observed (or maybe they have been!)---that employ quite different sets of causal cones. And~the cones of these <<other systems>> could very well lie outside the null cones of special relativity, i.e.,~these systems could very well manifest superluminal signals. None of this would contradict our fundamental ideas about how physics is structured: An initial value formulation, causal cones governing signals, etc.'' 

\section{Final~Words}
As it is said in~\cite{73}: ``Science, particularly mathematics, though~it 
seems less practical and less real than the news contained in the latest radio dispatches, appears to be building the one permanent and stable edifice in an age where all others are either crumbling or being blown to bits''
We firmly believe that special relativity belongs to this permanent and stable edifice.  Even if in the end it turns out that there are Lorentz non-invariant hidden sectors, special relativity will remain with us, as~one of our most precious scientific theories. Alas, this does not mean it will last forever, as~nothing lasts forever in the material world, especially such a fragile thing as intelligent life, and~particularly in context of current climate change. To~paraphrase English writer Virginia Woolf, it may happen that the very stone one kicks with one's boot will outlast all our theories and~achievements.

\vspace{6pt} 

\section*{Acknowledgments}
Shock and rage caused by Lucas Moodysson's film ``Lilja 4-ever'' triggered our musings on what last forever in our imperfect world and gave us an impetus to write this essay. Our special respects to Oksana Akinshina, a~Russian actress who played the main~role.

The work is supported by the Ministry of Education and Science of
the Russian~Federation.

\end{document}